\documentclass[conference]{IEEEtran}
\IEEEoverridecommandlockouts
\usepackage{cite}
\usepackage{amsmath,amssymb,amsfonts,hyperref,cleveref,booktabs,xcolor}
\usepackage{algorithmic}
\usepackage{graphicx}
\usepackage{textcomp}
\usepackage{xcolor}
\def\BibTeX{{\rm B\kern-.05em{\sc i\kern-.025em b}\kern-.08em
    T\kern-.1667em\lower.7ex\hbox{E}\kern-.125emX}}

\newcommand{\dis}[1]{\paragraph*{\textbf{Op#1}}}

\begin{document}

\title{An Exploration of Agile Methods in the Automotive Industry: Benefits, Challenges and Opportunities\\
}

\author{\IEEEauthorblockN{
Mehrnoosh Askarpour}
\IEEEauthorblockA{\textit{General Motors Canada} \\
Toronto, Canada \\
0000-0001-6526-2544}
\and
\IEEEauthorblockN{
Sahar Kokaly}
\IEEEauthorblockA{\textit{General Motors Canada} \\
Toronto, Canada \\
0009-0000-5348-2240}
\and
\IEEEauthorblockN{
Ramesh S}
\IEEEauthorblockA{\textit{General Motors R\&D} \\
Michigan, USA \\
0000-0002-8501-7447}

}

\maketitle

\begin{abstract}
Agile methodologies have gained significant traction in the software development industry, promising increased flexibility and responsiveness to changing requirements. However, their applicability to safety-critical systems, particularly in the automotive sector, remains a topic of debate. This paper examines the benefits and challenges of implementing agile methods in the automotive industry through a comprehensive review of relevant literature and case studies. 
Our findings highlight the potential advantages of agile approaches, such as improved collaboration and faster time-to-market, as well as the inherent challenges, including safety compliance and cultural resistance. By synthesizing existing research and practical insights, this paper aims to provide an understanding of the role of agile methods in shaping the future of automotive software development.
\end{abstract}
\section{Introduction}
The automotive industry is undergoing a profound transformation, driven by advancements in technology and changing consumer preferences\cite{cao2023agile}. In this dynamic landscape, the development of software-intensive systems plays a crucial role in enabling innovation and differentiation among automotive manufacturers. Agile methodologies have emerged as a popular approach to software development, emphasizing iterative development, customer collaboration, and adaptability to change. While agile methods have been widely adopted in various domains, their suitability for safety-critical systems, such as those found in advanced automobiles, presents unique challenges and opportunities. 

The adoption of agile methods in the automotive industry, such as the Scaled Agile Framework (SAFe)\cite{safe}, Large-Scale Scrum (LeSS)\cite{less}, or Disciplined Agile® Delivery (DAD) offers several potential benefits. Several articles in the literature discuss each of these frameworks with detail. For example,  Putta et al.\cite{DBLP:conf/esem/PuttaUHPL21} surveys practitioners motives to use SAFe;  Marinho et al.\cite{9503379} explores the use of SAFe in global software development; Edison et al.\cite{9387593} compares SAFe, LeSS, Scrum-at-Scale, and DAD; Almeida et al.\cite{almeida2021large} and Ozkan er al.\cite{DBLP:conf/enase/OzkanT20} provided a comprehensive study to enable organizations to select the framework that suits them best.
Irrespective of the specific framework used, the concept of agile practices has proven to be highly beneficial in the realm of software and system development. However, it is important to acknowledge that along with these benefits, the adoption of agile practices also presents new challenges.

This paper aims to explore the most common benefits and challenges of using agile methods in software intensive safety critical systems, drawing insights from recent studies, mostly interviews, conducted by researchers in the field. 
While previous studies by Alahyari et al.\cite{ALAHYARI2017271} and Heeager et al.\cite{HEEAGER201822} served as a foundation for more specific research, they were conducted in 2017 and 2018 respectively, and a comprehensive and updated study of the fast-paced industry is currently lacking. This paper can serve as a preliminary step towards conducting such a study.

It is important to note that not all of the cited papers in this study specifically focus on the automotive industry. Applying a domain-specific filter would have significantly reduced the number of papers included. This is particularly relevant considering our interest in examining papers published after 2015. However, the benefits and challenges discussed in this paper are applicable to all safety-critical domains, indicating their relevance beyond a specific industry. Indeed, domain-specific papers, if any, are project-specific and do not address the organization-wide values of agile methodologies.

The paper is structured as follows: \Cref{sec:benefits} highlights the most commonly reported benefits of using agile in the industry; \Cref{sec:challenges} features it's most commonly reported challenges; \Cref{sec:discussions} derives some open questions and potential research directions less explored in the literature; Finally, \Cref{sec:conc} concludes.

\section{Benefits of Agile Methods in the Automotive Industry}
\label{sec:benefits}
The implementation of agile methods in the automotive industry revolutionized the way projects are approached and executed. This section explores the specific advantages that agile methods offer to the automotive industry.

Firstly, most of the agile methodologies promote attention to a number of common concerns when it comes to developing safety critical systems. More specifically, agile methodologies focus on regular safety analysis and auditing, incremental change, viewing safety as a shared responsibility between all of the stakeholders, and most importantly improving traceability\cite{10.1007/978-3-030-35333-9_26}. Consequently, living traceability pushes to implement automated or semi-automated tools that facilitate the creation, maintenance, and utilization of trace links. Such tools could significantly ease  to visualize and explore the impact of change to safety\cite{9108244}. 

Secondly, according to Islam et al.\cite{ISLAM2020106954} who conducted several interviews with industry representatives using agile methods, in addition to enhanced engagement and communication, agile methods contribute to earlier discovery of problems in the software. 

Moreover, in addition to all the benefits aforementioned, Kasauli et al.\cite{8498249} performed a literature review that describes extra benefits for agile methods including efﬁcient use of available information, improved safety culture, prioritisation, quality, reuse of frameworks or work-products, reduced costs, and better test-cases.

\Cref{tab:benefits} summarizes the benefits of agile methods that we found in the literature through a snowballing approach focusing on articles that discuss industrial case-studies.

\begin{table*}[ht]
    \centering
    {\scriptsize
    \begin{tabular}{|p{0.1\linewidth}|p{0.4\linewidth}|p{0.08\linewidth}|p{0.18\linewidth}|p{0.07\linewidth}|}
\toprule
 \textbf{Benefit} & \textbf{Description} & \textbf{Papers} & \textbf{Paper Types} & \textbf{Related Challenges} \\\toprule
 Regular safety analysis \& auditing &  To prioritize safety in design decisions, programming, and validation, it is seamlessly integrated into the process and consistently addressed throughout rather than as isolated tasks. The organization aligns its process with customer standards or safety protocols along the development process.&\cite{10.1007/978-3-030-35333-9_26,8491141,9108244,myklebust2016agile,10.1145/3120459.3120482,myklebust2018agile} &interview, case study literature review, case study, case study, review& a,b,c,e\\\midrule
 Incremental change &The ability to incrementally update the safety cases based on small changes and the incorporation of variability in the safety analysis are further pre-requisites to achieve organisational flexibility.&\cite{10.1007/978-3-030-35333-9_26,myklebust2016agile,9dc411ecd9be454c8ab00718b6813a8a,9736397}&interview, case study, tool, interview& a,b,c,e\\\midrule
 Shared responsibility &Developers are never absolved from taking responsibility for safety in the ongoing safety activities. They need to be able to work with the risk analysis, update hazard logs and other artefacts, and ensure in their design decisions that safety considerations are upheld&\cite{10.1007/978-3-030-35333-9_26,8498249,10.1145/3120459.3120482,DBLP:journals/isj/VenkateshTCHS20,9736397}&interview,interview and literature review, case study, theory,interview& b,d \\\midrule
 Traceability&As the basis of testing, transparency, and smooth communication, traceability is required to ensure requirements are met and to define the needed testing.&\cite{10.1007/978-3-030-35333-9_26,9108244,8498249,10.1145/3387904.3389251,8029969}&interview, tool, interview and literature review, literature review, tool& a,e\\\midrule
 Earlier discovery of problems and risk anticipation& By continuously collecting feedback, issues can be identified and addressed early on, avoiding costly rework that often occurs during the integration phase. &\cite{ISLAM2020106954,linden2023scaling,KASAULI2021110851}&interview, case study & c\\\midrule
 Optimal information utilization &The organization prioritizes adaptability and responsiveness to change over rigidly following a plan, enabling flexibility and iterative development.&\cite{8498249}& interview and literature review & b\\\midrule
 Improved safety culture &Prioritizing safety from the early stages enhances safety awareness within the team and among stakeholders, ultimately fostering a stronger safety culture.&\cite{8498249}&interview and literature review & a,b,c,d,e\\\midrule
 Improved prioritisation&By prioritizing the highest value features, the organization ensures a more comprehensive definition and validation of important (e.g., higher ASIL) requirements.&\cite{8498249,ALAHYARI201978,10.1145/3494885.3494912,JARZEBOWICZ20203446,abdelazim2020framework,solinski2016prioritizing,9736397}&interview and literature review, review,interview,tool, review,interview&b\\\midrule
 Improved quality and better test cases &Prioritization, early problem discovery, and traceability would lead to rigorous and targeted testing, which consequently improves quality. &\cite{8498249,8054890,10.1007/978-3-319-69926-4_48, 7945719,rahy2022managing}&interview and literature review, tool,case study, case study,literature review&a, c, d,e\\\midrule
 Reduced costs and time &Prioritization, early problem discovery, and traceability, and Agile`s simplicity principal leads to higher efficiency and better resource usage.&\cite{8498249,DUFFIE2017273,ALAHYARI201978}&interview and literature review, case study,interview&a, b, c, d,e\\\midrule
 Framework/work-product reuse&The organization creates reusable frameworks that can be utilized in future phases or projects.&\cite{8498249,ALAHYARI201978,myklebust2016agile,10.1145/3120459.3120482,8029969,McCarey2005RascalAR,pang2022reuse}&interview and literature review, interview, case study, case study, tool, tool, theory&a,e\\\midrule
 Continuous learning &The organization fosters ongoing knowledge enhancement, improved competence and  performance. It also invests in continuous improvement and innovation.&\cite{ISLAM2020106954,linden2023scaling,ANNOSI2020554}& interview, case study, case study&b, c, d\\\midrule
    \end{tabular}
    }
    \caption{Summarized the benefits of agile methods as explored in academic literature focusing on industrial experiences.}
    \label{tab:benefits}
\end{table*}

\section{Challenges of Implementing Agile Methods in the Automotive Industry}
\label{sec:challenges}
Despite the potential benefits, implementing agile methods in the automotive industry is not without challenges. Many of the benefits mentioned in \Cref{sec:benefits} are achievable only when agile methods are adopted effectively. In practice, however, it takes time for different organizations to implement and practice agile right by the book and often there are several challenges\footnote{Please note the challenges discussed in this context pertain to the practical implementation of agile methodologies, based on information gathered from interviews and case studies. These challenges do not reflect any inherent flaws or shortcomings of the agile concept itself.}.
Most of the articles mention in \Cref{tab:benefits}, are divided in two parts discussing benefits and challenges of agile methods. We provide a summary of our findings below.

\paragraph{Documentation}Kasauli et al.\cite{8498249} discovered that the six large Swedish product development companies that were interviewed, found it difficult to comply with domain standards through an agile method. Notander et al.\cite{10.1007/978-3-642-39259-7_23} argues that safety standards are typically prescriptive and are often described with a waterfall-based process in mind. This means that they provide specific requirements and guidelines to be followed in a sequential and linear manner, aligning with the traditional waterfall development approach.
Authors count also for human factors (skills, experience, and attitudes) in correct or faulty adoption of agile methods resulting in issues like lack of documentation. Additionally, Anjum and Wolff\cite{9337750} describe that agile
processes and V model from ISO26262 is still a developing area and needs to be further explored. 

Moreover, it is not easy for teams to balance the focus on documentation in standards with iterative and fast paced agile techniques\cite{Wang2017ASO}. As a result, some of the benefits of \Cref{tab:benefits} such as traceability, reuse, and optimal information use are not generated. 
More recently, tool-based solutions are introduced to automate document generation and make it easy for teams. For example, Cardoso Rodrigues et al.\cite{9985199} came up with a tool that has been used and verified in the aerospace area; Authors of \cite{heeager2020meshing} propose a rather theoretical solution to mesh agile solutions and document intensive safety verification process that shall follow safety standards; Rahy and M.Bass\cite{rahy2022managing} propose two new artefacts, Documentation Work Item and Safety Critical Work Item, as ways to manage tasks (similar to Jira tickets but with a predefined workflow and definition of done), to adjust the handling of documentation and safety critical requirements in an agile framework. These solution, although all validated and proven helpful, are yet to be comprehensive enough to be commercialized or widely used.

\paragraph{Upfront planning} Similar to the previous challenge, safety standards resemble waterfall approaches and upfront planning rather than iterative mindset. Moreover, practitioners of safety standards are much more acquainted with waterfall mindset. Agile teams are typically small teams that focus on everyday product quality and are accustomed to fast requirement changes. However, safety teams favor safety over everyday quality. Additionally, every requirement change entails thorough analysis of its safety implications\cite{maqsood2022agile}.
This issue is extensively discussed and broken down to smaller sub challenges in \cite{ISLAM2020106954} where some empirical solutions from their interviewees are reported as well. It continues to be one of the major difficulties that organizations face with. 

\paragraph{Identifying and engaging the right stakeholders} In large cooperates and organizations with many cross-functional teams, it might sometimes be confusing to identify and communicate relevantly to everyone. Moreover, safety experts are accustomed to communication through documents (e.g., safety case, criticality analysis). The reduced emphasis on documentation in agile methods may result in unclear safety documentation, which can impede effective communication among various stakeholders. It is important to find a balance between agile practices and the need for clear and comprehensive safety documentation so not to loose benefits such as regular safety analysis and auditing, or early risk discovery. This challenge is one that could be resolved over time as the teams get more accustomed to the adopted agile method in their organization. Stegh{\"o}fer et al.\cite{10.1007/978-3-030-35333-9_26} suggests modular safety cases an one possible solution to tackle this problem, so that multiple teams could focus on a part of it and attend the planning and update meetings that are related to their part.\\
Furthermore, even if the appropriate individuals are engaged at the correct moment, bottlenecks can still occur as a result of the Product Owner's responsibility to oversee a diverse array of matters encompassing both operational and strategic aspects of a product\cite{ozkan2020evaluation}.

\paragraph{Balancing out safety and flexibility} If the shared responsibility on safety is not well understood throughout the organization, development and safety teams would struggle to coordinate and keep track of changes in each iteration. Koopman uses an interesting analogy in his lectures\cite{Koop} to highlight the importance of following defined processes n by agile teams. He describes undefined and undisciplined processes as cowboy coding.  
Likewise, Kasauli et al.  \cite{8498249} calls attention to the importance of collective code ownership, having experts within the team, and ensuring team members are familiar with safety standards. These factors contribute to effective collaboration, expertise, and adherence to safety requirements in software development projects.
Moreover, the agile principles such as continuous learning, collaborative spirit, and retrospectives shall be taken seriously by the teams and the leadership. 
According to Stegh{\"o}fer et al.\cite{10.1007/978-3-030-35333-9_26}, flexibility needs to be achieved in three areas: ecosystems of components used with suppliers, change management within the organization, and working with critical artifacts. This includes adapting components, managing changes effectively, and being flexible in handling critical documentation and specifications. Overall, everyone need to be be attentive that safety is not lost among all the fast paced and frequent changes.

\paragraph{Requirement management} Requirement management, subsequent of challenge (a), has garnered significant attention from the academic community. Rasheed et al.\cite{rasheed2021requirement} conducted a thorough examination of agile requirement engineering and identified key challenges. These challenges include developers neglecting non-functional requirements (such as safety), frequent changes in requirements, customers struggling to articulate user stories, and issues with missing, conflicting, or ambiguous requirements.
In their research, Jarzebowicz and Weichbroth\cite{9371679} conducted a literature review and interviewed 10 experienced practitioners to understand the underlying causes of these challenges. They discovered that non-functional requirements continue to emerge at different stages of the project life cycle. The most popular methods for requirement elicitation are interviews and meetings with stakeholders, and technical stories are commonly used to document non-functional requirements.
To address these problems, Rasheed et al. \cite{rasheed2021requirement} propose the use of a change management process and the adoption of formal requirement models. Examples of existing formal requirement models include RE-KOMBINE\cite{10.1007/978-3-642-31095-9_25}, TECHNE\cite{jureta2010techne}, REKB\cite{6051656}, Simulink Requirements Table blocks\cite{simu}, FORM-L and Modelica\cite{bouskela2022formal}, conceptual user story models based on UML notation\cite{9945932}, and tabular modeling\cite{systems11070352}, among others. However, it is important to note that some of these suggestions may not be scalable for the realistic demands of the automotive industry\cite{ULUDAG2022111473} or may require extensive training in areas such as UML semantics and propositional logic.

\section{Opportunities}
\label{sec:discussions}
This section raises thought-provoking questions and highlights open research areas and opportunities that can benefit safety-critical software, including the automotive industry and cyber-physical systems communities.

\dis{1}Numerous research papers extensively explore agile methods for safety-critical systems through interviews with industry representatives. However, these interviews often lack crucial details about the context of safety-critical projects and the identities of the companies involved. Consequently, categorizing the most prevalent benefits and challenges across industries becomes difficult. \textbf{Conducting a dedicated study on the unique challenges faced by the automotive software industry would greatly contribute to enhancing their software processes, ensuring customer safety, and mitigating the risks of expensive and reputation-damaging recalls.}

\dis{2}Based on the findings presented in Sections II and III, it is evident that the challenges and benefits identified in interviews regarding agile methods in safety-critical systems exhibit similarities. This suggests that the implementation of agile approaches in such systems is still in its nascent phase. Many of these challenges persist without definitive resolutions, leading companies to develop individualized, makeshift solutions. \textbf{To address this, establishing a knowledge base of best practices would facilitate the sharing of these solutions, even if they are proprietary. }Furthermore, this knowledge base would assist in data collection and contribute to the development of comprehensive solutions that could potentially be integrated into existing frameworks like SAFe.

\dis{3}According to Notander et al.\cite{10.1007/978-3-642-39259-7_23}, human factors play a crucial role in the successful adoption of an agile mindset, both in development and management. Hennel et al.\cite{hennel2021investigating} examine the impact of a healthy and supportive organizational environment, specifically psychological safety, on the performance of agile teams. The state of practice in agile automotive published by kUGLER's report\cite{kUGLER} highlights psychological safety and micro-management as two important existing challenge in the industry.\\
Surprisingly, none of the papers we reviewed considered this factor seriously. None of them included questions about human factors in their interviews with industry representatives. While they acknowledged the importance of teamwork and continuous learning as benefits, they did not measure or discuss these factors when addressing the challenges faced by interviewees. \textbf{The importance of soft skills in effectively managing an agile team is an intriguing area that has received limited attention and should be further explored.}\\
Although not extensively discussed in the literature (we only found it in \cite{8498249}), the lack of management support and trust in agile practices has been identified as a significant challenge. This was reported by approximately 40\% of participants in kUGLER's report\cite{kUGLER}. These findings highlight the importance of giving more attention to human factors and their impact on agile implementation.

\dis{4} As mentioned in challenge (c), incorporating modular safety cases and regular meetings, such as PI plannings and stand-ups, can assist in merging agile practices with regulatory constraints. However, these solutions still tend to be ad hoc and implemented on a case-by-case basis. It would be valuable to identify specific issues that arise when propagating safety cases in agile projects and develop strategic and comprehensive solutions for them. This endeavor could ultimately result in the creation of automation tools that facilitate the integration of safety engineers into agile teams. \textbf{Furthermore, this pursuit could lead to the formulation of more precise and improved instructions for modular safety cases within safety and regulatory standards.}

\dis{5} The academic literature on agile has steadily grown since as early as 2001, indicating a strong interest and exploration within the research community. However, there has been a decline in the number of studies conducted since 2019. While this may suggest that the topic is perceived as mature and no longer requires further investigation, the industry, particularly the automotive sector, continues to face challenges, particularly in the areas of verification, validation, and the integration of software development, system engineering, and safety. This is evident from recent recalls by major companies (e.g.,\cite{tesla1,tesla2,gm1,toyota1,fiat1}), who are still striving to align these aspects while introducing new features like over-the-air updates. \textbf{Therefore, it is our belief that the industry still greatly benefits from the innovative and exploratory thinking of the academic community. Their contributions are essential in classifying and disseminating valuable best practices, as well as inventing and automating methods for traceability, modular safety case development, documentation, and more.}

\dis{6} The compatibility and scalability of Model Based Engineering (MBE) with agile approaches have sparked ongoing debate \cite{mukaienvisioning,DBLP:journals/jss/LiebelK23}; Guerrero{-}Ulloa et al\cite{DBLP:journals/sensors/GuerreroUlloaRH23} even argue that they represent distinct development methodologies. However, other researchers advocate for further exploration of MBE and agile \cite{DBLP:journals/sosym/BurguenoCFKLMPP19,DBLP:conf/modelsward/FantP19,miller}. 

Holtmann et al\cite{DBLP:journals/jss/HoltmannLS24} conducted an extensive literature review and interviews to assess the alignment between agile development and MBE. Their conclusion suggests that the two approaches do not hinder each other, although integrating their respective tools remains a significant challenge. \textbf{Hence, we see an opportunity for deeper exploration of the relationship between these two concepts. It's essential to channel research and industrial efforts into developing tools and approaches that facilitate the integration of both methodologies.}

\section{Conclusions}
\label{sec:conc}

The adoption of agile methodologies in the automotive industry represents both a significant opportunity and a formidable challenge. As highlighted throughout this paper, agile methods offer a plethora of benefits, including enhanced safety practices, improved communication and collaboration, early problem discovery, and cost reduction. These advantages are instrumental in driving innovation and competitiveness in an industry undergoing rapid technological transformation.

However, the implementation of agile practices in safety-critical systems presents several hurdles that organizations must overcome. These challenges include issues related to documentation, upfront planning, stakeholder engagement, requirement engineering, and the delicate balance between safety and flexibility. Addressing these challenges requires a concerted effort from industry practitioners, researchers, and policymakers alike.

Moreover, there are several open questions and research opportunities that merit further exploration. These include the need for more industry-specific studies to understand the unique challenges faced by the automotive sector, the establishment of a knowledge base of best practices to facilitate learning and knowledge sharing, and a deeper investigation into the role of human factors in agile implementation.

Despite the growth in academic literature on this topic, there remains a continued need for innovative thinking and exploration. The automotive industry, in particular, continues to grapple with complex issues such as verification, validation, and the integration of software development with system engineering and safety. In this context, the contributions of the academic community are invaluable in driving progress and ensuring the safe and efficient development of software-intensive systems in the automotive industry and beyond.

\bibliographystyle{IEEEtran}
\bibliography{bib.bib}
\end{document}